\documentclass[a4paper,11pt]{article}
\pdfoutput=1

\usepackage{jheppub}
\usepackage{amsmath,amssymb,amsfonts}
\usepackage{graphicx}
\usepackage{cleveref}
\usepackage{hyperref}
\usepackage{srcltx}
\usepackage{bm}
\usepackage{mathdots}
\usepackage{color}
\usepackage{dsfont}
\usepackage{slashed}
\usepackage{physics}
\usepackage{subfigure}
\usepackage{tikz}
\usepackage[normalem]{ulem}
\usepackage{comment}

\newcommand{\bea}{\begin{eqnarray}}
\newcommand{\eea}{\end{eqnarray}}
\newcommand{\beq}{\begin{equation}}
\newcommand{\eeq}{\end{equation}}
\newcommand{\mbaryo}{m_{\mathcal{B}_{\rm SM}}}
\newcommand{\mpsi}{m_{\psi_\mathcal{B}}}
\newcommand{\psiB}{\psi_\mathcal{B}}
\newcommand{\tabBR}{\mathrm{Br}(B\!\to\! \mathcal{B}_\mathrm{SM}\! +\! \psi_\mathcal{B})}

\newcommand{\FR}{$F^{\mathcal{O}}_{R}$}
\newcommand{\FL}{$\widetilde{F}^{\mathcal{O}}_{L}$}
\newcommand{\tabO}[1]{$\mathcal{O}=\mathcal{O}_{#1,u_i d_j}$}
\newcommand{\tabOb}[1]{$\mathcal{O}=\mathcal{O}_{#1,u_i b}$}
\newcommand{\LT}{$\widetilde{L}_1$}

\title{Branching Fractions of $B$ Meson Decays in Mesogenesis}
\preprint{MITP-22-096}
\author[a]{Gilly Elor,}\emailAdd{gelor@uni-mainz.de}
\author[b]{Alfredo Walter Mario Guerrera} \emailAdd{guerrera@pd.infn.it}
\affiliation[a]{\sl PRISMA$^+$ Cluster of Excellence \& Mainz Institute for Theoretical Physics\\
Johannes Gutenberg University, 55099 Mainz, Germany}
\affiliation[b]{INFN, Sezione di Padova}

\abstract{Production of the matter-antimatter asymmetry in the $B$-Mesogenesis mechanism is directly related to the branching fraction of seemingly baryon number violating decays of $B$ mesons into a light Standard Model baryon and missing energy. Achieving the observed baryon asymmetry requires that the branching fraction for such decays be greater than about $10^{-7}-10^{-5}$. Experimental searches at $B$ Factories and Hadron Colliders target specific decay modes. Therefore, computing the exclusive branching fraction for each decay is a critical step towards testing Mesogenesis. In this work we use QCD Light Cone Sum Rules to compute the form factors and branching fractions of the various possible channels contributing to the baryon asymmetry. Using the results, we comment on implications for current and future experimental searches. 
}

\begin{document}
\maketitle
\flushbottom
\setcounter{page}{2}
\newpage

\section{Introduction}
Under the standard paradigm of inflationary cosmology the Universe is born with equal parts matter and anti-matter. This necessitates a dynamical mechanism of \emph{baryogenesis} to generate the asymmetry necessary to seed the complex structures observed today. The required primordial baryon asymmetry  is inferred to be $Y_{\mathcal{B}}^{\rm meas} \equiv (n_{B}-n_{\bar{B}})/s = \left( 8.718 \pm 0.004 \right) \times 10^{-11} $
from measurements of the Cosmic Microwave Background (CMB)~\cite{Planck:2015fie,Planck:2018vyg} and light element abundances after Big Bang Nucleosynthesis (BBN)~\cite{Cyburt:2015mya}. 
Another outstanding mystery is the nature and origin of dark matter; the gravitationally inferred, but yet to be experimentally detected, component of matter which makes up roughly 26\% of the energy of the Universe today~\cite{Planck:2015fie,Planck:2018vyg}. 
Recently proposed mechanisms of \emph{Mesogenesis}~\cite{Elor:2018twp,Elor:2020tkc,Elahi:2021jia} use Standard Model (SM) meson systems to generate both the baryon asymmetry and the dark matter abundance of the Universe.

Many mechanisms have been proposed since 1967 when Sakharov first introduced the  conditions~\cite{sakharov} necessary for baryogenesis;  C and CP Violation (CPV), baryon number violation, and departure from thermal equilibrium. Traditional mechanisms of baryogenesis often involve high scales and massive particles leading to dismal prospects for experimental verification\footnote{See~\cite{Elor:2022hpa} for a summary of some interesting new ideas for testing traditional mechanisms.}. In contrast,  mechanisms of Mesogenesis generate the baryon asymmetry of the Universe by leveraging the CPV within SM meson systems. Excitingly, Mesogenesis is directly testable at hadron colliders and $B$-factories~\cite{Alonso-Alvarez:2021qfd, Borsato:2021aum}, with additional indirect signals at Kaon and Hyperon factories~\cite{Alonso-Alvarez:2021oaj,Goudzovski:2022vbt}  and even upcoming neutrino experiments such as DUNE and Hyper-Kamiokande~\cite{MesoDUNE} (see also~\cite{Elor:2022hpa,Asadi:2022njl,Barrow:2022gsu,Fox:2022tzz} for summaries). A search for operator through which neutral $B$-Mesogenesis can proceed has already been conducted by the Belle-II collaboration~\cite{Belle:2021gmc}, and a search has also been proposed at LHCb~\cite{Rodriguez:2021urv}.

Production of the baryon asymmetry in the Neutral $B$-Mesogenesis mechanism~\cite{Elor:2018twp} is directly related to the branching fraction of seemingly baryon number violating decays of neutral $B_{s,d}^0$ mesons into a light SM baryon and missing energy.
Successful baryogenesis in this framework predicts this observable should be $\text{Br} ( B_{s,d}^0 \rightarrow \mathcal{B}^0_{\rm SM} + \text{MET} )  \gtrsim \mathcal{O}(10^{-5})$  to generate the observed baryon asymmetry~\cite{Elor:2018twp,Alonso-Alvarez:2021qfd}. Charged $B$ Mesogenesis~\cite{Elahi:2021jia}, links the baryon asymmetry to the decays of the $B^\pm$ mesons and requires $\text{Br} ( B^+ \rightarrow \mathcal{B}^+_{\rm SM} + \text{MET} )  \gtrsim \mathcal{O}(10^{-6})$. Furthermore, a measurement of the branching fraction as low as $10^{-7}$ could be an indication that a Mesogenesis mechanism is at play.

The nature of the final state SM baryon $\mathcal{B}_{\rm SM}$ in these decays depends on the flavor structure of the operator through which $B$-Mesogenesis proceeds (see Table~\ref{tab:decays} for a list of all the possible contributing decays). 
Experimental searches target $B$ meson decay modes to specific final states, for instance the Belle-II collaboration set a limit on the branching fraction for $B_d^0$ decaying into a neutral $\Lambda$ baryon and missing energy~\cite{Belle:2021gmc}. Therefore,  the  computation of 
the exclusive branching fractions of the various possible decay modes is critical for testing $B$-Mesogenesis. Searching for such $B$ meson decays down to the $\text{Br} \sim 10^{-7}$ level will fully test the landscape of different Mesogenesis possibilities. 

In this work we leverage the powerful machinery of QCD  Light Cone Sum Rules (LCSRs), first introduced
in~\cite{BALITSKY1989509,Braun:1988qv,Chernyak:1990ag}, reviewed in~\cite{Colangelo:2000dp},
to compute the branching fractions of the various channels that could generate the baryon 
asymmetry.  The hard scattering amplitude of the $B$ meson to baryon transition is convoluted with the Distribution Amplitude (DA) of the baryon; these are fundamental non perturbative functions that can be interpreted as light-cone wave functions
integrated over transverse quark momenta~\cite{Lepage:1980fj,RQCD:2019hps}. Using our results, we comment on implications for current and future experimental searches in light of 
existing collider and flavor constraints on the model. 

This paper is organized as follows: in Sec.~\ref{sec:review} we first review the mechanism of $B$-Mesogenesis. We specifically emphasize the details most useful for an experimentalist eager to conduct such searches. Next, in Sec.~\ref{sec:LCSR}, we detail the results of the LCSRs derivation of the form factors and branching fraction of the relevant $B$ meson decays. 
Our main results are presented in Figure~\ref{fig:Br_light} and \ref{fig:Br_heavy}
and in Table~\ref{tab:decays}, we further discuss the results and implications for Mesogenesis in Sec.~\ref{sec:results}. We conclude in Sec.~\ref{sec:conc} with an outlook on future directions.
Appendices ~\ref{sec:FF_LCSR}, \ref{sec:charmed}, and \ref{app:Numeric} contain supplementary details on the formalism and derivation.

\section{Overview of $B$-Mesogenesis}
\label{sec:review}
Here we provide a brief review of the neutral $B$-Mesogenesis mechanism, with particular emphasis on the those ingredients most relevant for experimental searches. 

The baryon asymmetry and dark matter abundance are produced through  $B$-Mesogenesis  as follows: we first assume the late time out of equilibrium production of equal numbers of $B_{s,d}^0$, $\bar{B}_{s,d}^0$ and $B^\pm$ mesons\footnote{This can be achieved if a scalar field, perhaps related to inflation, comes to dominate the energy density of the Universe at late times.} when the temperature of the Universe was $5 \,\text{MeV} \lesssim T \lesssim 100 \, \text{MeV}$ i.e. after the QCD phase transition but before BBN. The produced $B$ mesons will then undergo SM CP violating processes before decaying into a dark sector state charged under SM baryon number and a SM baryon. The end result is an equal and opposite baryon asymmetry in the dark and the visible sector which is directly proportional to the branching fraction of the exotic  decay of the $B$ mesons. In the case of Neutral $B$-Mesogenesis~\cite{Elor:2018twp}, the generated asymmetry will also depend on the charge asymmetry  of the $\bar{B}^0-B^0$ oscillations, and indirectly on a host of other observables (see~\cite{Alonso-Alvarez:2021qfd} for an overview). 
Achieving the observed baryon asymmetry~\cite{Alonso-Alvarez:2021qfd}, in light of current world averages limiting the CPV in the meson systems, requires an inclusive branching fraction $\text{Br} ( B^0_{s, d} \rightarrow   \psiB + \mathcal{B}_{\rm SM}) \gtrsim 10^{-5}$. Where $\psiB$ is a dark fermion carrying baryon number $-1$, and $\mathcal{B}_{\rm SM}$ is a neutral SM baryon. Charged $B$ Mesogenesis~\cite{Elahi:2021jia}, leverages CPV in SM meson decays and requires $\text{Br} ( B^+ \rightarrow \mathcal{B}^+_{\rm SM} + \psiB )  \gtrsim \mathcal{O}(10^{-6})$ to generate the observed baryon asymmetry\footnote{Note that fully verifying this \emph{flavor} of Mesogenesis requires future generation experiments, and as such is less testable on shorter time scales than its neutral counterpart.}. 
In both cases the exact nature of the SM baryon depends on the flavor structure of the UV model as will be discussed below. Indeed, various different decay modes can contribute to the generation of the baryon asymmetry.

\begin{figure}[t!]
    \begin{center}
        \includegraphics[width=1.0\textwidth]{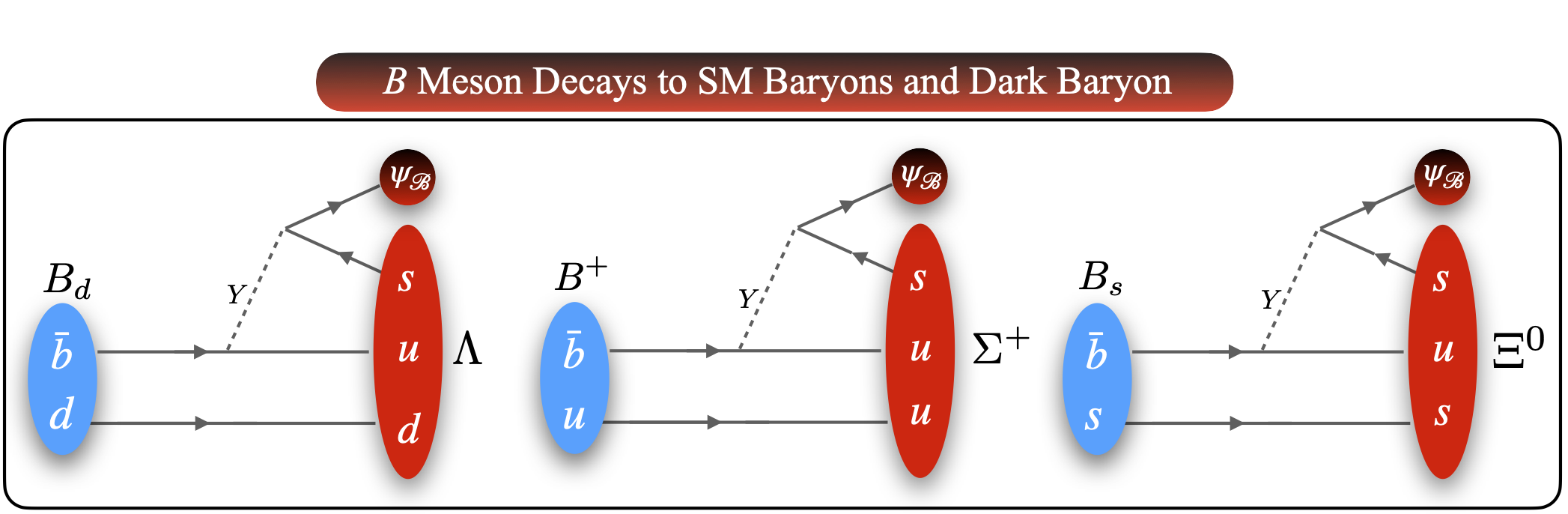}
    \end{center}
    \caption{Decay modes of  $B_{s,d}^0$ and $B^+$ mesons through the $\mathcal{O}_{b,us}$ and $\mathcal{O}_{s,ub}$  operators. Decays through the other operators listed in Table~\ref{tab:decays} arise in a similar way.}
    \label{fig:CartoonDecays}
\end{figure}

The \emph{seemingly} baryon number violating decay of the charged or neutral $B$ meson can generically proceed as follows: 
a colored scalar mediator\footnote{\footnotesize{ See~\cite{Alonso-Alvarez:2019fym} for a consistent UV realization of this theory where the heavy colored mediator is identified with a squark of a supersymmetric theory.}} $Y$ can mediate the decay of a $b$-quark within the $B$ meson to a SM baryon $\mathcal{B}_{\rm SM}$ and dark sector baryon $ \psi_{\mathcal{B}}$ (which appears as missing energy in the detector). The Lagrangian for the model, allowed by all the symmetries, is
\bea
\label{eq:Lmodel}
\mathcal{L} &=& -  \, y_{u_\alpha d_\beta} \epsilon_{ijk} Y^{*i}  \bar{u}_{\alpha}^j d_{\beta}^{c,k} - \, y_{\psi d_\gamma} Y_i \bar{\psi}_{\mathcal{B}} d_{\gamma}^{c,i}  + \text{h.c.}\,,
\eea
where all SM quarks are right handed, and ``c'' indicates a charge conjugation. $i,j,k$ are color indices, and $\alpha, \beta, \gamma$ are flavor indices. 
Here $Y$ has the SM charge assignment (3,1,-1/3) \footnote{Note that other charge assignments for the mediator are also possible e.g. $(3,1,2/3)$ or $(3,2,-1/6)$ which lead to a variety of different flavor and collider constraints~\cite{Alonso-Alvarez:2021qfd,Alonso-Alvarez:2021oaj}), and furthermore could allow the couplings of $\psiB$ to left handed quarks. For simplicity, we consider only the hypercharge $-1/3$ mediator case in the present work.}. 
The colored mediator $Y$ can be produced at the LHC and thus must have a mass at the TeV scale to be consistent with collider constraints~\cite{Alonso-Alvarez:2021qfd}.
Integrating out $Y$ leads to the following effective operator at the low energies:
\bea
\label{eq:OpBdecay}
&& \mathcal{H}_{d_\gamma, u_\alpha d_\beta}=- \frac{ y_{\psi d_\gamma}  y_{u_\alpha d_\beta}}{M_Y^2} i \epsilon_{ijk} \left( \bar{\psi}_\mathcal{B} d_{\gamma}^{c,i}\right)\left(  \bar{u}_{\alpha}^j d_{\beta}^{c,k} \right) + \text{h.c.}\,.
\eea
There are four different flavor combinations that correspond to operators that can mediate the decay of the $b$ quark within the $B^0$ meson and thus contribute to the generated baryon asymmetry in the $B$-Mesogenesis framework. Each flavor combination can give rise to two possible effective operators:
\bea
\mathcal{O}_{b,u_\alpha d_\beta } \equiv i \epsilon_{ijk} b^i \left( \bar{d}_\beta^{c,j} u_\alpha^k \right) \quad \text{and} \quad \mathcal{O}_{d_\beta,u_\alpha b } \equiv i \epsilon_{ijk} d_\beta^i \left( \bar{b}^{c,j} u_\alpha^k \right)\label{eq:OpsDef}
\eea
where again the quarks are right handed and $u_\alpha = u\,,c$ and $d_\beta = d\,, s$; giving eight operators in total. 

\begin{table}[t]
\renewcommand{\arraystretch}{1.2}
  \setlength{\arrayrulewidth}{.25mm}
\centering
\small
\setlength{\tabcolsep}{0.18 em}
\begin{tabular}{ |c || c | c | c | c | c | c |}
    \hline
      Flavorful 			&   Decay                     	                & \FR, \FL       &\FR, \FL           & $\tabBR$                &  $\tabBR$                  \\ 
  \,\,\, Operator \,\,      &   Channel            	                        &\tabO{b}        &\tabOb{d_k}         & \tabO{b}               & \tabOb{d_k}          \\ \hline
        \hline
					  	    &   $B_d\to \psi_\mathcal{B} + n$           	&$R_1,$ 0   &$R_1,$ 0 &$3.7_{\pm 0.4}\!\cdot\!10^{-7}$&$8.3_{\pm 0.9}\!\cdot\!10^{-6}$ \\ 
$\,\, \psi_\mathcal{B} \, b\, u\, d \,\,$ 	&   	$B_s\to \psi_\mathcal{B} + \Lambda$            & n.a.	           & n.a.	        &          n.a              &        n.a.                   \\ 
                            &   	$B^+\to \psi_\mathcal{B} + p$	                     &$R_2$, \LT &$R_1,$ 0  &$9.6_{\pm 0.6}\!\cdot\!10^{-8}$& $8.9_{\pm 0.9}\!\cdot\!10^{-6}$       \\ \hline\hline 
						    &   	$B_d\to\psi_\mathcal{B} + \Lambda$                &$R_1,$ 0   &$R_1,$ 0  &$1.2_{\pm 0.06}\!\cdot\!10^{-5}$ &$3.2_{\pm 0.6}\!\cdot\!10^{-5}$       \\ 
$ \psi_\mathcal{B} \, b\, u\, s$        &   	$B_s\to\psi_\mathcal{B} + \Xi^0$       	        &$R_2$, \LT &$R_3$, \LT &$2.6_{\pm 0.1}\!\cdot\!10^{-6}$&$4.8_{\pm 0.2}\!\cdot\!10^{-5}$\\ 
                            &   	$B^+\to\psi_\mathcal{B} + \Sigma^+$            &$R_2$, \LT &$R_1,$ 0   &$2.5_{\pm 0.3}\!\cdot\!10^{-6}$&$2.1_{\pm 0.2}\!\cdot\!10^{-5}$\\  \hline\hline 
						    &   	$B_d\to \psi_\mathcal{B} + \Sigma_c^0$          &$R_2$, \LT &$R_3$, \LT  &$1.3_{\pm 0.6}\!\cdot\!10^{-6}$&$2.7_{\pm 1.5}\!\cdot\!10^{-4}$\\ 
$\psi_\mathcal{B} \, b\, c\, d$        &   	$B_s\to\psi_\mathcal{B} + \Xi_c^0$	           &$R_2$, \LT &$R_3$, \LT  &$1.2_{\pm 0.6}\!\cdot\!10^{-6}$&$2.3_{\pm 1.3}\!\cdot\!10^{-4}$\\ 
                        	&       $B^+\to\psi_\mathcal{B} + \Sigma_c^+$          &$R_2$, \LT &$R_3$, \LT&$1.4_{\pm 0.7}\!\cdot\!10^{-6}$&$2.9_{\pm 1.6}\!\cdot\!10^{-4}$\\  \hline\hline
						    &   	$B_d\to\psi_\mathcal{B} + \Xi_c^0$		        &$R_2$, \LT &$R_3$, \LT&$8.4_{\pm 3.7}\!\cdot\!10^{-6}$&$3.2_{\pm 1.7}\!\cdot\!10^{-4}$\\ 
$\psi_\mathcal{B} \, b\, c\, s$       &   	    $B_s\to\psi_\mathcal{B} + \Omega_c$              &$R_2$, \LT &$R_3$, \LT&$9.0_{\pm 4.0}\!\cdot\!10^{-6}$&$4.4_{\pm 2.0}\!\cdot\!10^{-4}$\\ 
                           &     	$B^+\to\psi_\mathcal{B} + \Xi^+_c$		        &$R_2$, \LT &$R_3$, \LT&$1.8_{\pm 0.6}\!\cdot\!10^{-5}$&$3.5_{\pm 1.9}\!\cdot\!10^{-4}$\\   \hline

\end{tabular}
\caption{ We summarize the possible $B$ meson decay channels which can generate the baryon asymmetry in $B$-Mesogenesis.
For each operator and decay mode we list the resulting structure of the form factor $R_{1,2,3}$ or $\tilde{L}_1$ given in Eqs.~(\ref{eq:R1}--\ref{eq:L1}). In the final two columns we quote the maximum possible branching fraction for each operator. These are found by evaluating the branching fraction at $\mpsi = 1$ GeV, and by fixing the Wilson coefficient to its maximum possible value allowed by LHC constraints, computed in~\cite{Alonso-Alvarez:2021qfd}. Note that in the case of $B_d \rightarrow \psiB + \Sigma_c^0$, the very prompt decay $\Sigma_c^0 \rightarrow \Lambda_c^+ + \pi^-$ will result in additional subtitles for experimental reconstruction.
}
\label{tab:decays}
\end{table}

In Table~\ref{tab:decays} we list the four possible flavorful variations of the operator in
Eq.~\eqref{eq:OpsDef} along with the corresponding $B$ meson decay modes, through which the baryon
asymmetry can be generated can proceed. Note that the model also allows the decay of $B^+$ mesons
which, while not directly responsible for the generated baryon asymmetry, serves as an indirect
probe of the mechanism. In Figure~\ref{fig:CartoonDecays} we show the decays through the
$\mathcal{O}_{b,us}$/$\mathcal{O}_{s,ub}$  operator for the $B_{s,d}^0$ meson as well as the $B^+$
meson. Note that the coefficients $y_{\psi d_\gamma}  y_{u_\alpha d_\beta}/M_Y^2 $ associated with
the above operators are constrained by a combination of LHC searches and flavor observables 
(see~\cite{Alonso-Alvarez:2021qfd} for a detailed analysis). For simplicity, where obvious, 
we will omit the flavor indices on the Wilson coefficient and use refer to it as $y^2/M_Y^2$.

Paramount to confirming Mesogenesis as the mechanism by which Nature chose to generate the baryon asymmetry is the measurement of the various decay modes in Table~\ref{tab:decays}. This is the main result of the present work.

Some final comments on the parameter space are now in order. The mass of the dark baryon $\psi_\mathcal{B}$, and thus the corresponding missing energy, is required to be within a window of roughly 1-4 GeV. The reason is as follows: First it must be the case that $m_{\rm \psi_\mathcal{B}} < m_B - m_p \simeq 4.34 \, \text{GeV}$, 
such that the requisite decay of the $B$ meson is kinematically allowed. Second, since the operator Eq.~\eqref{eq:OpBdecay} can lead to proton decay, we simply require that any dark sector state charged under baryon number be sufficiently heavy to kinematically forbid such decays and therefore stabilize matter\footnote{Neutron star bounds may place slightly more stringent constraints on the mass of dark baryons (see~\cite{McKeen:2018xwc}) but these are model dependent so we can allow ourselves to be more relaxed.}: $m_{\psi_\mathcal{B}} > m_p - m_e \simeq 937.8 \text{MeV} $. 
In summary, the dark baryon should have a mass in the range
\bea
m_p - m_e < \mpsi < m_B - \mbaryo \,,
\eea
corresponding to a range about 1-4 GeV. 

Finally, note that $\psi_\mathcal{B}$ cannot be the dark matter as it is unstable. Indeed, the operators of Eq.~\eqref{eq:OpBdecay} would allow the decay $\psi_\mathcal{B} \rightarrow p +  e + \bar{\nu}_e$ which is kinematically allowed, and would washout the generated asymmetry. To evade washout and stabilize the dark matter additional dark sector states must be introduced. The dark sector dynamics has no implications on the current work and we simply refer the interested reader to~\cite{Elor:2018twp,Elahi:2021jia} for more details. 

\section{Form Factors from Light Cone Sum Rules}
\label{sec:LCSR}

Historically, the LCSR technique was developed to explain the mass difference 
between the $\Sigma$--hyperion and the proton in the weak decay $\Sigma\to p + \gamma$~\cite{BALITSKY1989509}. Technically, LCSR combines the QCD sum rules technique  with the theory of hard 
exclusive processes. 
Intuitively, the short distance expansion is replaced by the
expansion in the transverse distance between partons in light-cone coordinates
~\cite{Braun:1997kw}.
This allows one to incorporate
additional information about QCD correlators related to the approximate conformal 
symmetry of the theory. The order of the expansion is regulated by the \emph{twist} of the operators~\cite{OHRNDORF198226,Braun:1989iv}. 
Twist is defined as the difference between the dimension and the spin of an operator; increasing the twist accuracy takes into account higher transverse momentum and higher Fock state contributions~\cite{Ball:1998fj}.

To compute the branching fractions of the two body meson to baryon decays listed in Table~\ref{tab:decays} via the QCD LCSR technique, one introduces a baryon to vacuum correlator. This object depends on the specific meson interpolating 
current and on the effective three-quark operator coupled to $\psi_\mathcal{B}$. In the recent paper~\cite{Khodjamirian:2022vta}\footnote{\cite{Khodjamirian:2022vta} appeared
while the present work was being prepared.} the two body decay $B^+\to p+\psi_\mathcal{B}$
was obtained using the LCSR method. Our approach reproduces the result of~\cite{Khodjamirian:2022vta},
and we further generalize it to all the remaining decays that could contribute to the production of the baryon asymmetry. Unless explicitly stated otherwise, we will mostly
refer to the $B$--meson triplet, $(B_d,B_s,B^+)$, without specifying the quark content, simply as $B$. 
A brief summary of the formalism is given in Appendix \ref{sec:FF_LCSR}, and we extended it to heavy baryons in Appendix \ref{sec:charmed}.

The two body decay amplitudes of interest all have the form: 
%
\begin{equation}
    \mathcal{A}(B\to \mathcal{B}_{SM}+\psi_\mathcal{B})= \frac{y_{\psi d_\gamma}  y_{u_\alpha d_\beta}}{M_Y^2} \bra{\mathcal{B}_{SM}(P)}
    \mathcal{O}\ket{B(P+q)}u^c_{\psi_\mathcal{B}}(q)\,,
    \label{eq:decay_amplitude}
\end{equation}
%
where $u^c_{\psi_\mathcal{B}}$ is the charge conjugated bispinor of the $\psi_\mathcal{B}$ field. Here $P+q$
and $P$ are the momenta of the $B$-meson and $\mathcal{B}_{SM}$ respectively while
 $y_{\psi d_\gamma} y_{u_\alpha d_\beta}/M_Y^2$ is the Wilson coefficient associated with the specific operator $\mathcal{O}$ defined in Eq.~(\ref{eq:OpBdecay}).
Lorentz 
covariance dictates that the hadronic matrix element in Eq.~(\ref{eq:decay_amplitude})
can be parameterized in terms of four independent form factors as follows:
%
\beq
    \bra{\mathcal{B}_{SM}(P)}\mathcal{O}\ket{B(P+q)}  =  F_R^{\mathcal{O}}(q^2)\bar{u}_{\mathcal{B}_R} + F_L^{\mathcal{O}}(q^2)\bar{u}_{\mathcal{B}_L}+\widetilde{F}_R^{\mathcal{O}}(q^2)\frac{\slashed{q}}{\mbaryo}\bar{u}_{\mathcal{B}_R} + \widetilde{F}_L^{\mathcal{O}}(q^2)\frac{\slashed{q}}{\mbaryo}\bar{u}_{\mathcal{B}_L},
    \label{eq:form_factors_decomposition}
\eeq
%
where $\bar{u}_{\mathcal{B}_{R,L}}$ is the chiral spinor associated with the baryon.
The calculation of the various $F_I$'s appearing in Eq.~(\ref{eq:form_factors_decomposition})
is analogous to the ones discussed in~\cite{Khodjamirian:2022vta}, and we refer the interested reader to the details in Appendix~\ref{sec:FF_LCSR}. Of these four possible structures only two are found to be non-zero for the specific 
choice of transition operators we consider, namely, $F_R$ and $\tilde{F}_L$ ---  a consequence of taking only
$R-$handed fields in Eq.~(\ref{eq:OpsDef}). The different flavor structures produce three distinct forms of $F_R$
and a single form $\tilde{F}_L$. 
All the transitions in Table~\ref{tab:decays} are calculated using the leading twist-3 DAs: 
%
\bea
    R_1(q^2) & = & \frac{m_b^3}{4 m_B^2 f_B}\int_0^{\alpha_0^B}
    \!\!\! d\alpha \,\,\,e^{(m_B^2-s(\alpha))/M^2}\!\left\{ \frac{\widetilde{V}(\alpha)}{(1-\alpha)^2}\left(1+\frac{(1-\alpha)^2 \mbaryo^2-q^2}{m_b^2}\right)\right\}\!,\label{eq:R1}\\
    R_2(q^2) & = & \frac{m_b \mbaryo^2}{4 m_B^2 f_B}\int_0^{\alpha_0^B}
    \!\!\! d\alpha \,\,\,e^{(m_B^2-s(\alpha))/M^2}\left( \widetilde{V}(\alpha)-3 \frac{m_b}{\mbaryo}\frac{\widetilde{T}(\alpha)}{(1-\alpha) }\right),\label{eq:R2}\\
    R_3(q^2) & = & \frac{m_b^3}{4 m_B^2 f_B}\int_0^{\alpha_0^B}
    \!\!\! d\alpha \,\,\,e^{(m_B^2-s(\alpha))/M^2}\!\left\{ \left(1-\frac{q^2}{m_b^2}\right)\frac{\widetilde{V}(\alpha)}{(1-\alpha)^2}+3\, \frac{\mbaryo \widetilde{T}(\alpha)}{\,\,m_b\,\, (1-\alpha) }\right\}\!,\label{eq:R3}\\
    \widetilde{L}_1(q^2) & = & \frac{m_b \mbaryo^2}{4 m_B^2 f_B}\int_0^{\alpha_0^B}
    \!\!\! d\alpha \,\,\,\frac{\widetilde{V}(\alpha)}{(1-\alpha)}\,\,e^{(m_B^2-s(\alpha))/M^2}\label{eq:L1}. 
\eea
%
Where in Eqs.~(\ref{eq:R1}--\ref{eq:L1}) $s(\alpha)=[m_b^2-\alpha q^2+\alpha(1-\alpha)\mbaryo^2]/(1-\alpha)$, and
\beq
\alpha_0^B=\frac{s_0^B-q^2+\mbaryo^2-\sqrt{(s_0^B-q^2+\mbaryo^2)^2 - 4\mbaryo^2(s_0^B-m_b^2)}}{2 \mbaryo^2}\,.
\eeq
Here $s_0^B$ and $M^2$ are the effective threshold and the Borel 
parameter respectively. They correspond to ``internal" parameters
of the LCSR description that are to be fixed.
The two functions appearing in the definitions of the sum rules above
are the, once integrated, leading twist-3 DAs of for the specific baryon, given by~\cite{RQCD:2019hps}: 
%
\beq
\widetilde{V}(\alpha)=\int_0^{1-\alpha}\!\!\!\! d \alpha_2\,\, [V_1+A_1]{(\alpha,\alpha_2,1-\alpha-\alpha_2)}\,\,,
\eeq
%
%
\beq
\widetilde{T}(\alpha)=\int_0^{1-\alpha}\!\!\!\! d \alpha_2\,\, T_1(\alpha,\alpha_2,1-\alpha-\alpha_2),
\eeq
%
The functions $V_1$, $A_1$ and $T_1$ are defined in Eqs.~(1-9) of~\cite{RQCD:2019hps} or in Eqs.~(6-9) of~\cite{Anikin:2013aka}. 
The definitions of the sum rules for charmed operators are not modified 
however subtleties are involved in the DAs of heavy--baryons; a dedicated discussion is given in \ref{sec:charmed}.
From  Eqs.~(\ref{eq:decay_amplitude}--\ref{eq:form_factors_decomposition})
one can recover a convenient expression for the two-body branching ratio mediated by a single operator:
\bea
    \mathrm{Br}(B\,\!\!\to\!\!\,\mathcal{B}_\mathrm{SM}&\!\!\!+&\!\!\psi_\mathcal{B}) \!=\! \frac{|y_{\psi d_\gamma}  y_{u_\alpha d_\beta}|^2\tau_B}{32 \pi\, m^3_B M_Y^4}\!\left\{\left|R_I(m_{\psi_\mathcal{B}}^2)\!+\!\frac{m_{\psi_\mathcal{B}}}{\mbaryo}\widetilde{L}_I(m_{\psi_\mathcal{B}}^2)\right|^2\!\!\!(m_B^2\!-\!\left(\mbaryo\!-\!m_{\psi_\mathcal{B}})^2\right) \right.\nonumber\\
     &+& \left. \left|R_I(\mpsi^2)\!-\!\frac{\mpsi}{\mbaryo}\widetilde{L}_I(\mpsi^2)\right|^2\!\!\!\!\left(m_B^2\!-\!(\mbaryo+\mpsi)^2 \right) \!\!\right\}\lambda^{1/2}(m_B^2,\mbaryo^2,\mpsi^2)\,,\nonumber\\
    \label{eq:Br}
\eea
where $\lambda$ is the K\"{a}llen function. 
The appropriate $R_I$ and $\widetilde{L}_I$ can be read off the corresponding entry in Table~\ref{tab:decays}.

\section{Results}
\label{sec:results}

In this work we have used LCSRs to compute the exclusive branching fractions for each decay mode in Table~\ref{tab:decays} that could generate the baryon asymmetry in the Mesogenesis framework. Our results for the \emph{maximum possible branching fraction} (in light of LHC constraints) are presented in Figure~\ref{fig:Br_light} and  Figure~\ref{fig:Br_heavy}  for $B$ meson decays to light and heavy (charmed) baryons respectively. In what follows we will discuss some interesting features of these results, before moving on to comment on implications for testing $B$-Mesogenesis.

\subsection{Discussion of Results}
\label{subsec:DisResults}
\begin{figure*}[t]
\centering
\setlength{\tabcolsep}{12pt}
\renewcommand{\arraystretch}{1}
\begin{tabular}{cc}
		\label{fig:BrObud}
		\includegraphics[width=0.45\textwidth]{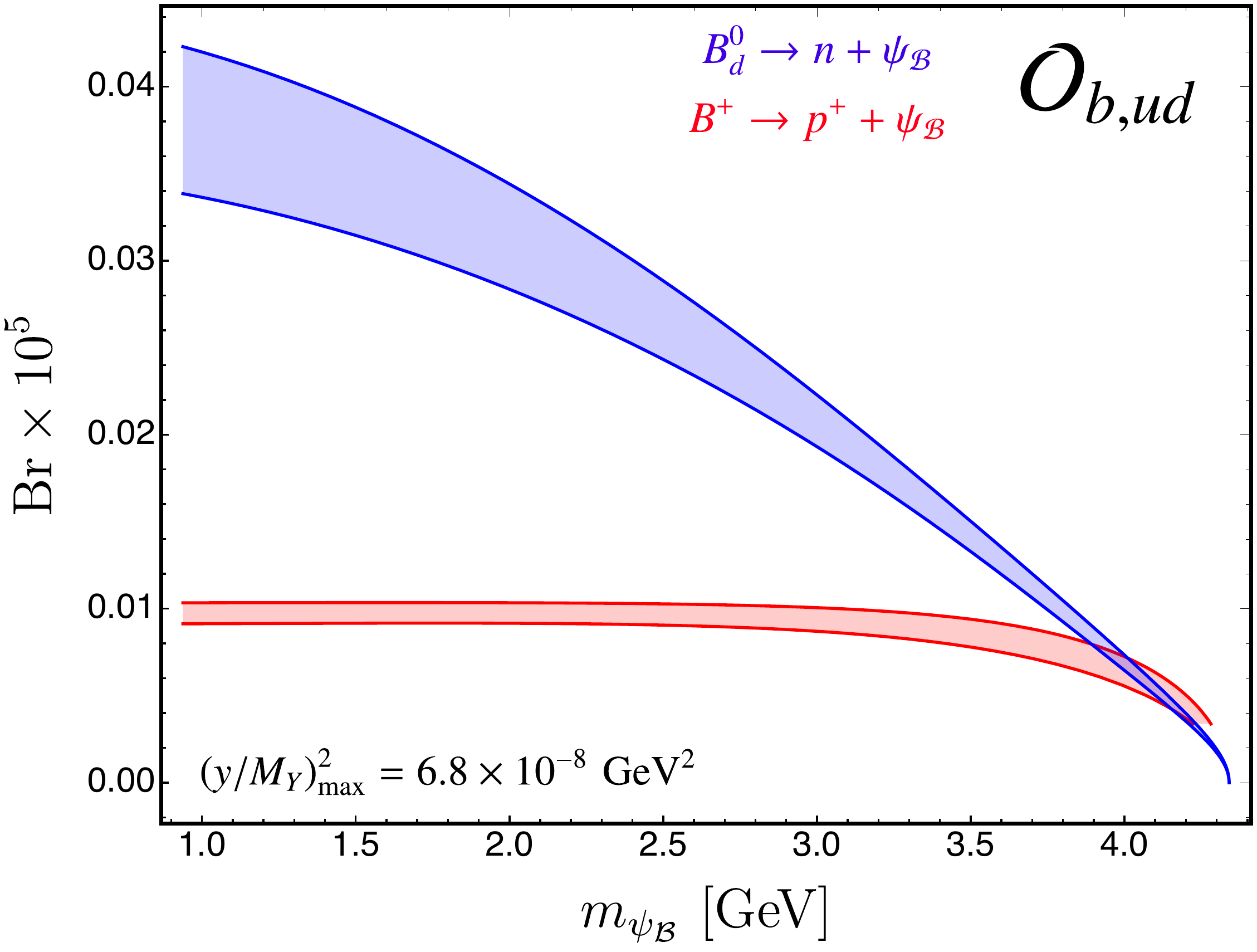}
&		
		\label{fig:BrOdub}
		\includegraphics[width=0.44\textwidth]{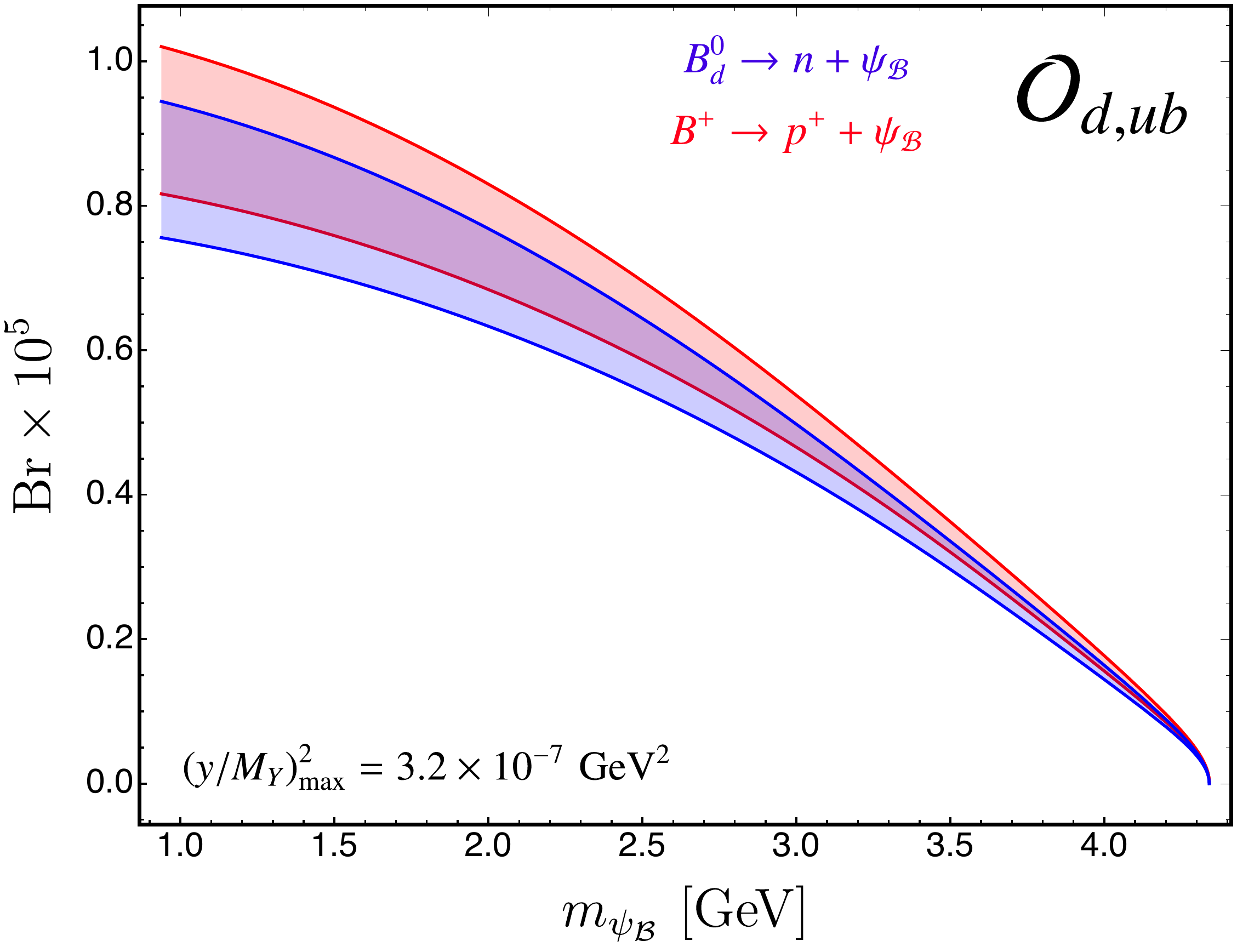} \\
		
    	\label{fig:BrObus}
		\includegraphics[width=0.45\textwidth]{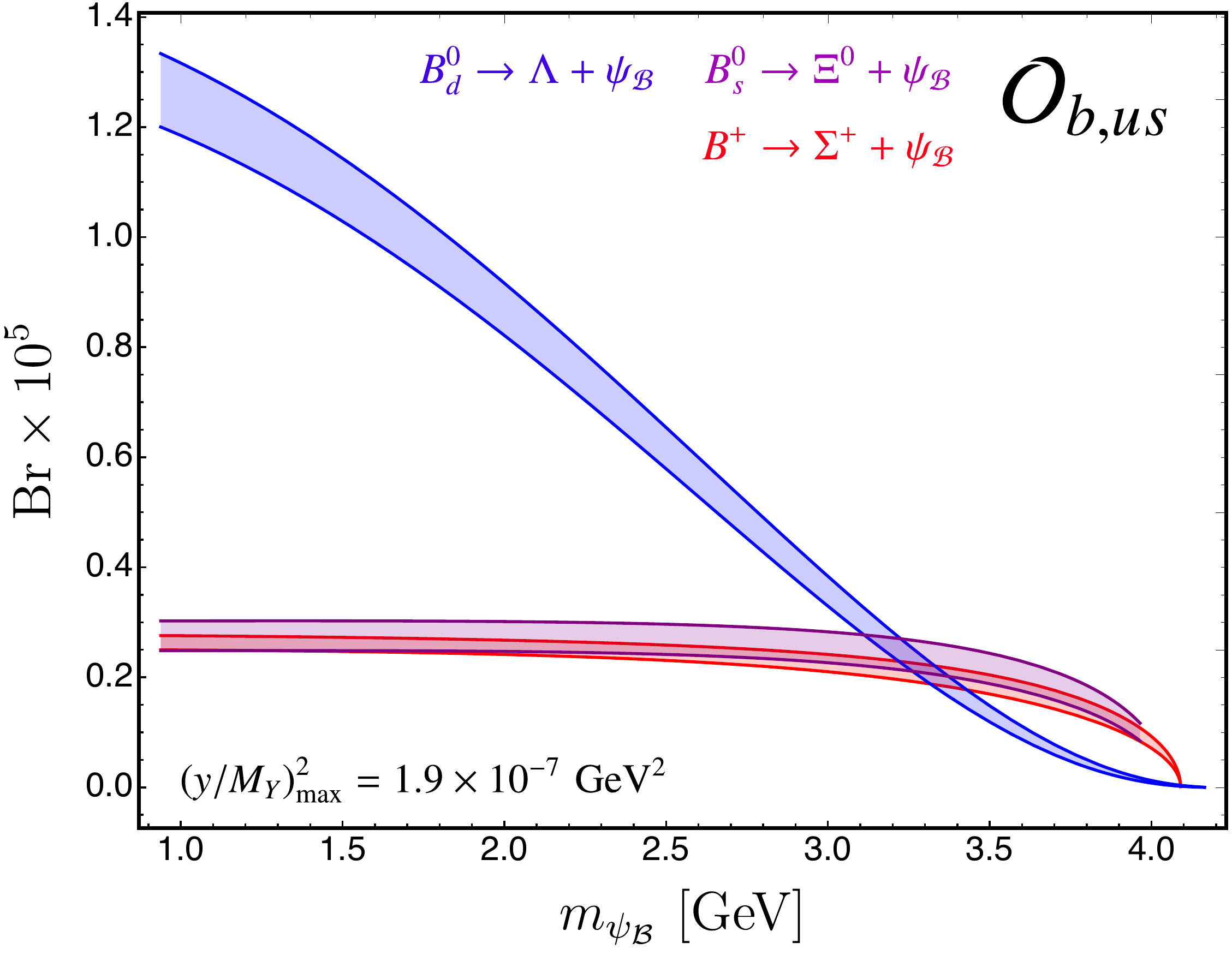}
&		
		\label{fig:BrOsub}
		\includegraphics[width=0.44\textwidth]{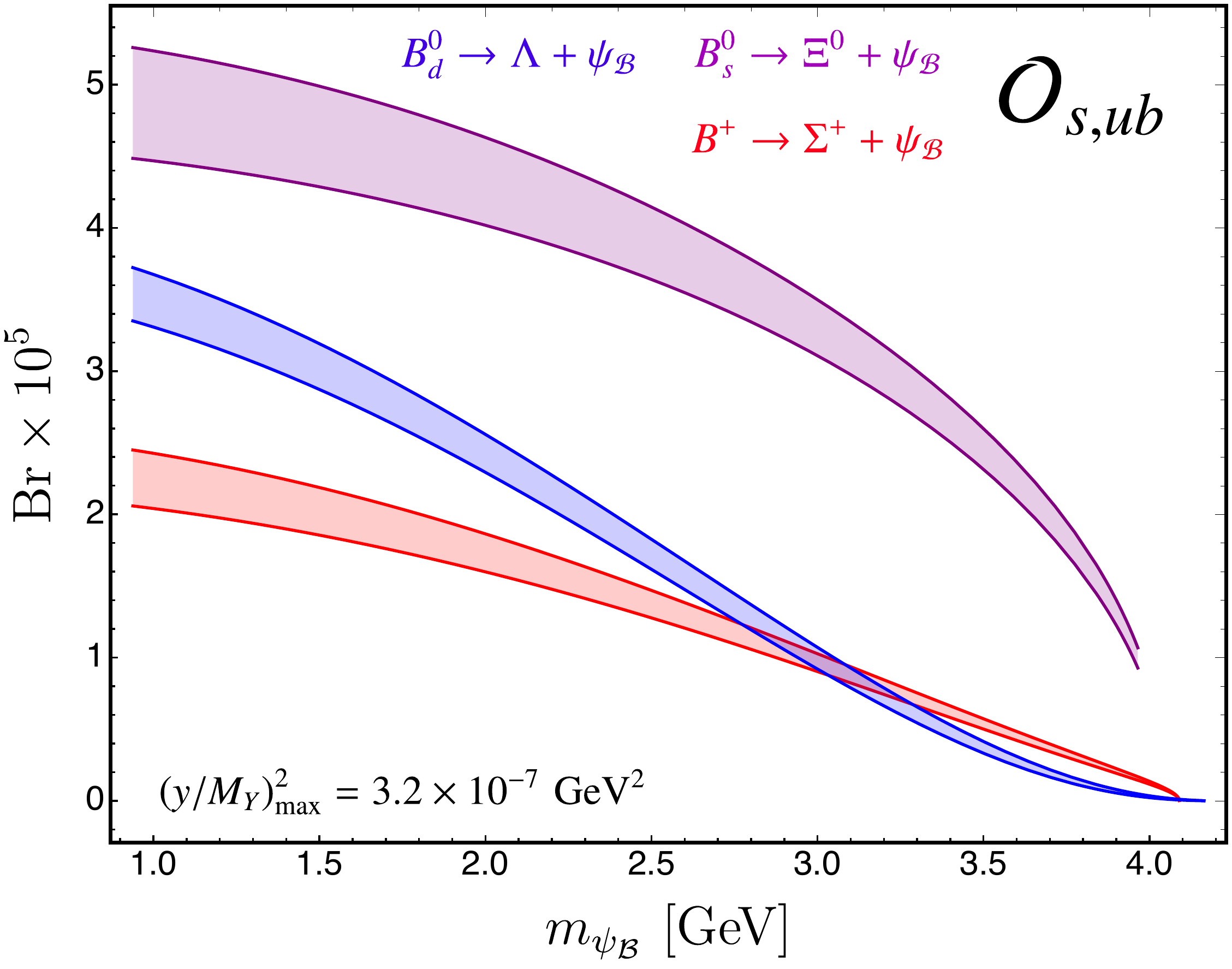}
		
\end{tabular}
\vspace{-0.2cm}
\caption{ Branching fractions as a function of $\psiB$ mass for the decays arising from the light quark operators $\psiB \,b\,u\,d$ and $\psiB \,b\,u\,s$; the top two rows of Table~\ref{tab:decays}.
For each operator, we have fixed the Wilson coefficient to the maximum possible value allowed by LHC constraints, computed in~\cite{Alonso-Alvarez:2021qfd}. 
The error bands come primarily from the uncertainties in the DAs and the errors on the internal parameters, as discussed in the text. }
\label{fig:Br_light}
\end{figure*}

\begin{figure*}[t]
\centering
\setlength{\tabcolsep}{12pt}
\renewcommand{\arraystretch}{1}
\begin{tabular}{cc}
		\label{fig:BrObcd}
		\includegraphics[width=0.45\textwidth]{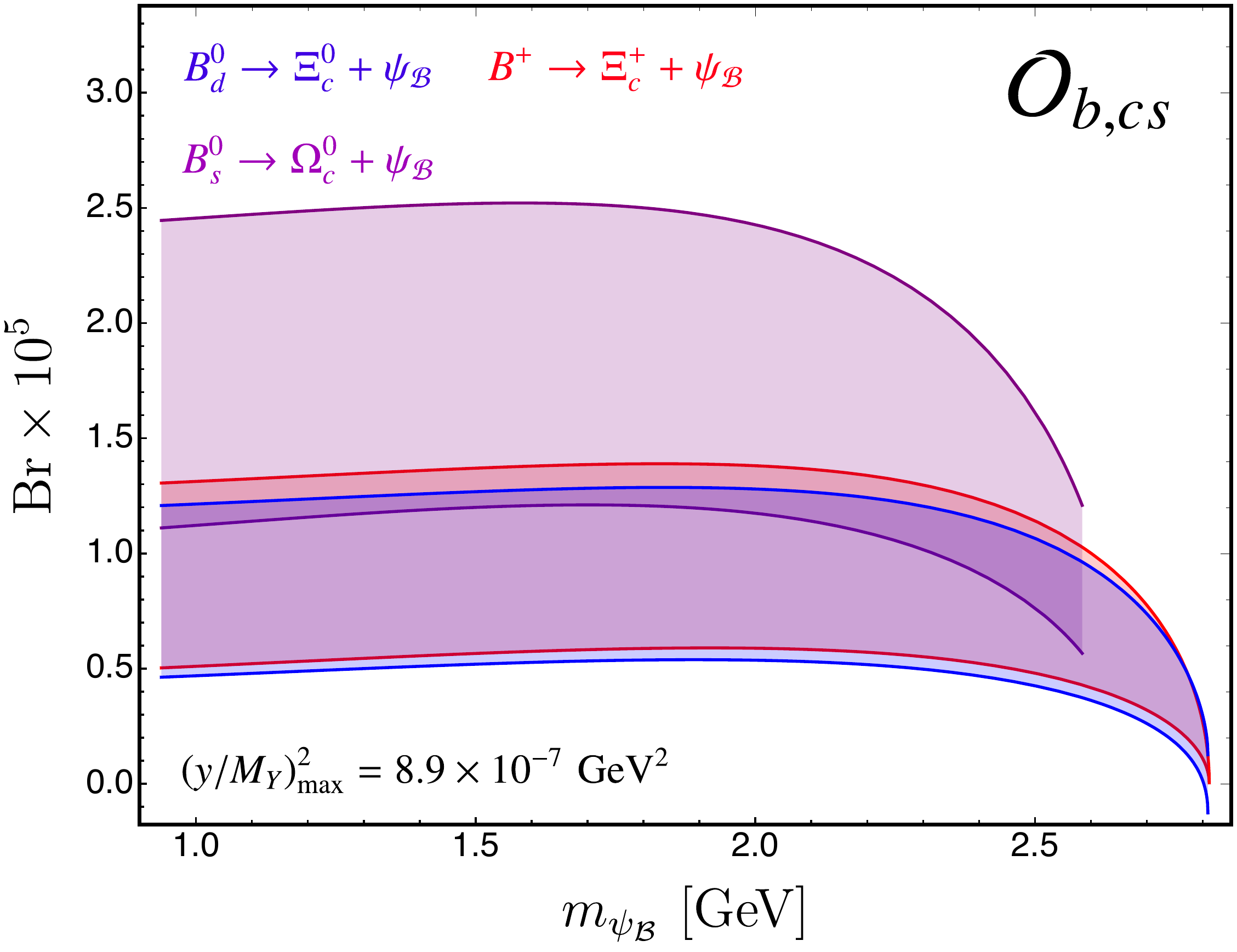}
&		
		\label{fig:BrOdcb}
		\includegraphics[width=0.44\textwidth]{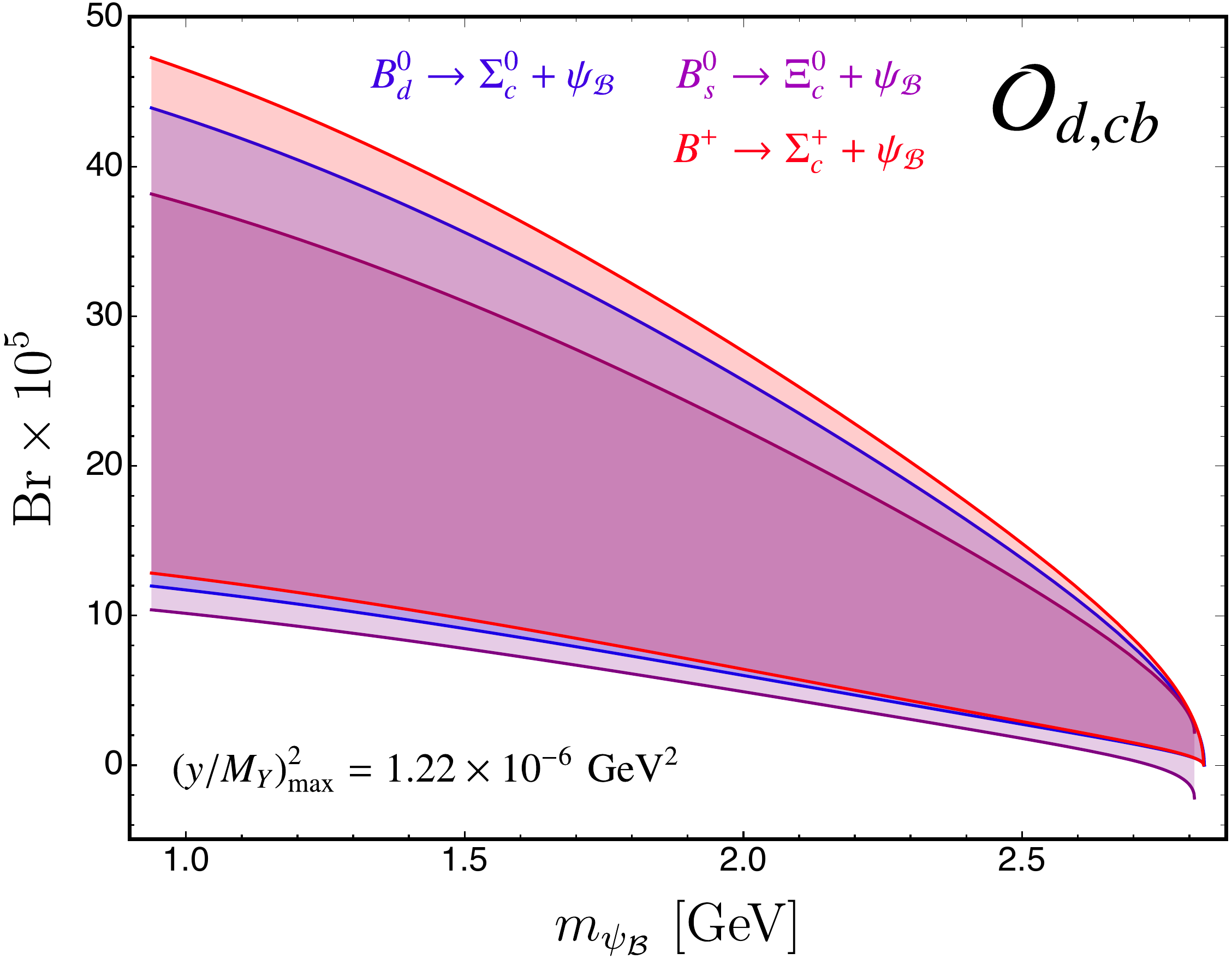} \\
		
		\label{fig:BrObcs}
		\includegraphics[width=0.45\textwidth]{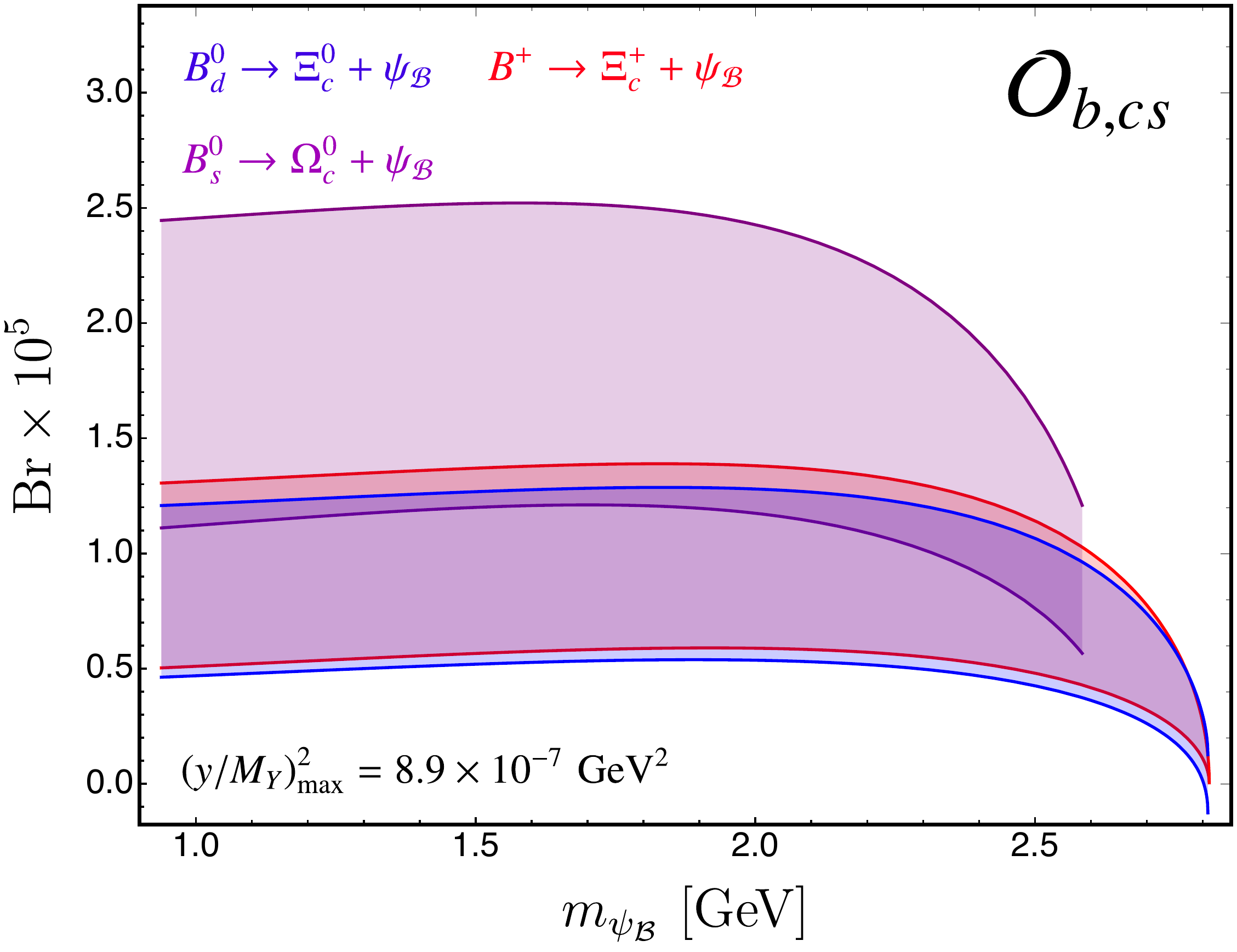}
&		
		\label{fig:BrOscb}
		\includegraphics[width=0.45\textwidth]{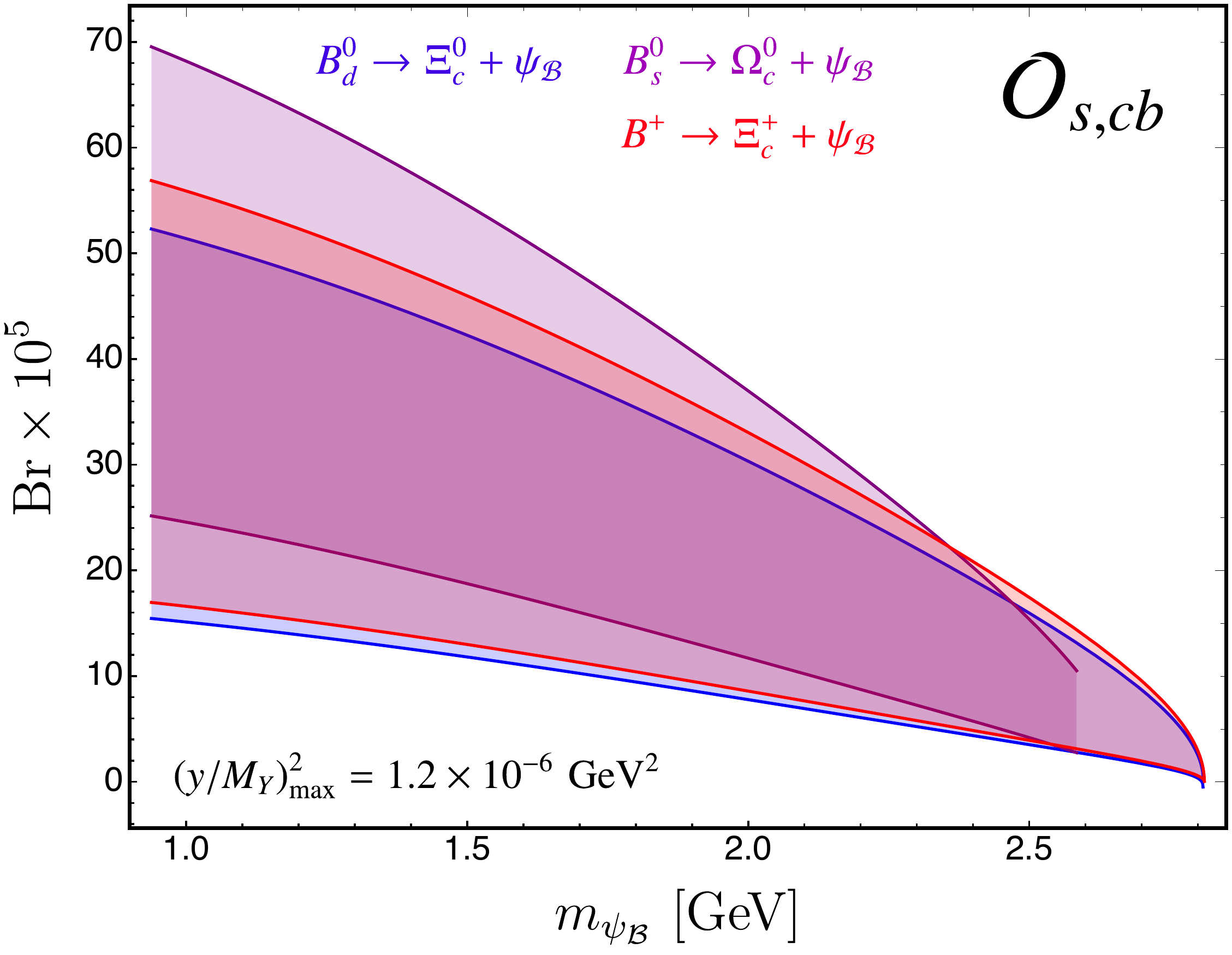}
		
\end{tabular}
\vspace{-0.2cm}
\caption{Branching fractions for decays from the charmed operators $\psiB \,b\,c\,d$ and $\psiB \,b\,c\,s$ ; the bottom two rows of Table~\ref{tab:decays}. See App \ref{sec:charmed} for a treatment of the DA for heavy quarks. For each operator, we have fixed the Wilson coefficient to the maximum possible value allowed by LHC constraints, computed in~\cite{Alonso-Alvarez:2021qfd}. The larger uncertainties come from an additional 15\% error that we apply on all parameters
entering the DAs definition inferred by heavy-quark symmetry}
\label{fig:Br_heavy}
\end{figure*}

By substituting the results obtained
in Appendix \ref{app:Numeric} for the form factors\footnote{In 
the following we will not explicitly distinguish form factors from
their fits with the BCL expansion. The reader should keep in mind that 
any numerical results will be obtained via the values of the fits.} 
 into Eq.~(\ref{eq:Br}), we obtain the exclusive branching for all processes shown in Table~\ref{tab:decays}. In Figure~\ref{fig:Br_light} the results for the maximum branching fraction from the first two flavorful operators, corresponding to decays of the $B$ meson to lighter baryons, are shown as a function of $\mpsi$.
The plots are ordered 
with $\mathcal{O}_{b,u_i d_j}$ on the left and $\mathcal{O}_{d_k,u_i b}$ on the right.
The first row corresponds to the operator  $\psiB \,b\,u\,d$, while the second to
$\psiB \,b\,u\,s$, mimicking the order of the top two rows of Table~\ref{tab:decays}. LHC searches for new colored particles limit the size of the  Wilson coefficient $(y/M_Y)^2$ for each operator --- these bounds were computed in~\cite{Alonso-Alvarez:2021qfd}. We fix $(y/M_Y)^2_\mathrm{max}$ so that the plotted branching fractions represent the maximum possible ones. To obtain a branching ratio independent on the Wilson coefficient one can simply re-scale the curves by a factor of $1/(y/M_Y)^4$. In Figure~\ref{fig:Br_heavy} we present the maximum branching fractions (again using LHC bounds on the specific operators) for decays to charmed baryons mediated by the operators $\psiB \,b\,c\,d$, and $\psiB \,b\,c\,s$ i.e. the two bottom rows of Table~\ref{tab:decays}. These branching fractions have been computed using the DAs from HQET by employing 
heavy-quark symmetry\footnote{A detailed discussion on charmed DAs can be found in Appendix \ref{sec:charmed}.}. The maximal values of $(y/M_Y)^2_\mathrm{max}$ are reported in each panel of Figure~\ref{fig:Br_light} and Figure~\ref{fig:Br_heavy}.
These are also used to compute the branching fractions values at $\mpsi=1$ GeV which are presented in the two right most columns of Table~\ref{tab:decays}. These entries represent the overall maximum possible branching fractions and are important for understanding the implications for Mesogenesis, as we will discus further below.

We can already learn a lot by considering the first operator 
$\psi_\mathcal{B} \,b\, u\, d$. For example it is interesting to see that 
$SU(2)_F$ violating effects are mainly due
to the masses entering in the branching ratio formula Eq.~\eqref{eq:Br},
as the form factors are not really affected by these differences 
(see Table~\ref{tab:fit}).
This can be seen in the fifth column of Table~\ref{tab:decays} as the entries for the $B_d$ and $B^+$
decay differ more than their respective form factors values. A second emerging pattern
is the sub-dominance of the $R_2$ and $\widetilde{L}_1$ structures in comparison to $R_1$ and $R_3$. This effect is seen in all the transitions except 
$B_d\to \psi_\mathcal{B} +n$. This suppression is simply due to the fact that the prefactors in Eq.~(\ref{eq:R1}) and (\ref{eq:R3})
are a factor of $(m_b/\mbaryo)^2$ larger than Eq.~(\ref{eq:R2}) and (\ref{eq:L1}).
The effect is even more pronounced in the heavy-baryon sector at small 
$\psi_\mathcal{B}$ masses as the contribution of $\widetilde{L}_1$ receives an extra suppression factor of  $m_{\psi_\mathcal{B}}/\mbaryo$. 
Therefore at higher values of $m_{\psi_\mathcal{B}}$ it is possible to see a small enhancement
in the decays mediated by $\mathcal{O}_{b,u_i d_k}$.
In general the $\mathcal{O}_{d_k,u_i b}$ operators will give
the higher contribution among the two.

Another notable feature in Table~\ref{tab:decays} is that the
$B_s \to \psi_\mathcal{B}+\Lambda$ transition appears to be prohibited at leading twist. 
The vanishing of the amplitude does not depends on the particular 
operator in Eq.~(\ref{eq:OpsDef}) nor on the particular chirality 
of its fields. 
This selection rule is due to conservation of total angular 
momentum. The bounded di-quark system in the $B_s$ meson will have total
spin $j=0.$ In the final state the spinless di-quark system is given by
the valence couple $(u,d)$ while the spin of the baryon is carried solely 
by the spectator $s-$quark~\cite{RQCD:2019hps}.
Therefore, conservation of total angular momentum imposes the spin of $\psi_\mathcal{B}$
to be opposite to the $s$ quark spin; this is forbidden by the scalar nature of the
operators studied in this work.

The maximum branching fractions for the heavier baryon final states in Figure~\ref{fig:Br_heavy} are typically an order of magnitude larger than the ones in Figure~\ref{fig:Br_light} simply due to the weaker LHC constraints on the charmed channels. 
The general features discussed for the first two operators still apply,
e.g. the hierarchy between $\mathcal{O}_{b,u_i d_j}$ and $\mathcal{O}_{d_k,u_i b}$ is still present. 

Finally, the uncertainties appearing in Figure \ref{fig:Br_light} are obtained by treating the internal 
parameters ranges as errors, (see App.~\ref{app:Numeric}), and adding them in quadrature with the errors of the DAs parameters. 
In Figure~\ref{fig:Br_heavy}, we note that the uncertainties are far more significant for the charmed baryon final states due to an additional 15\% error that we apply on all parameters entering the DAs definition inferred by heavy-quark symmetry~\cite{Khodjamirian:2017zvq}. Otherwise, uncertainties are propagated in the same way as for decays to light baryons.

\subsection{Implications for Mesogenesis}

With the results for the branching fractions in hand, we now comment on the implications for testing Mesogenesis.

The Neutral $B$ Mesogenesis~\cite{Elor:2018twp} mechanism is currently the most testable of all the Mesogenesis scenarios. As such, for practical reasons alone, it is motivated for experimentalits to first focus on targeting its parameter space. As discussed above, successful Neutral $B$ Mesogenesis requires an inclusive branching fraction that is greater than roughly $\text{Br}( B_{s,d}^0 \rightarrow \psiB + \mathcal{B}_{\rm SM}) \gtrsim 10^{-5}$, in light of existing bounds on the CPV in the $B^0-\bar{B}^0$ oscillation system. Several or all the $B_{s,d}^0$ decay modes listed in Table~\ref{tab:decays} could contribute to the generation of the asymmetry. Therefore, to \emph{fully test} Neutral $B$ Mesogenesis requires a \emph{search for all possible decays}. To date one such search has been conducted; 
the Belle-II collaboration~\cite{Belle:2021gmc}  preformed a search targeting the decays through the $\psiB \, b\, u\, s$ operator,  setting a limit $\text{Br} \left( B_d^0 \rightarrow \psiB + \Lambda \right) \lesssim 2 \times 10^{-5}$. Comparing this to the second row of Table~\ref{tab:decays} we see that this limit is \emph{not} more constraining than the theoretical predictions for the branching fraction in light of LHC bounds on the Wilson coefficient. Furthermore, examining the maximum branching fractions in the first row of Table \ref{tab:decays}, we see that it is only marginally possible for $\psiB \, b \, u \, d$ operator to be entirely responsible for the generation of the baryon asymmetry. Meanwhile, the other three operators have branching fractions that could still be large enough to be entirely reasonable for the baryon asymmetry. 

While Neutral $B$ Mesogenesis is exceptionally testable on a short time scale, it is critical to emphasize that this not the only possible way to generate the baryon asymmetry by leveraging SM Meson systems. Other mechanisms of Mesogenesis involve the CPV in the decays of Charged $D$ mesons~\cite{Elor:2020tkc}, or $B$ and $B_c$ mesons~\cite{Elahi:2021jia}. In particular, for $B_c$ Mesogenesis~\cite{Elahi:2021jia}, the exact same model Eq.~\eqref{eq:Lmodel} is evoked to generate the baryon asymmetry. In this case, it is the branching fraction of the $B^+$ decays that directly feeds into the generated baryon asymmetry. From Fig 2 of~\cite{Elahi:2021jia}, we can see that the branching fraction for $B^+ \rightarrow \psiB + \mathcal{B}_{\rm SM}$ could be as small as $10^{-6}$ and still reasonably generate the baryon asymmetry. As such we conclude that every one of the operators in Table \ref{tab:decays} is still a viable candidate for generating the asymmetry in this scenario. 

Allowing for the most general scenario, the branching fraction could be as low as $\text{Br} \gtrsim 10^{-7}$ and still generate the baryon asymmetry, if all the CPV came entirely from the dark sector. This scenario is far less reconstructable but is nevertheless worth keeping in mind when conducting searches. 

In summary, the branching fractions computed here provide experimentalists with a critical ingredient for testing Mesogenesis. Experiments such as Belle, BaBar, LHCb (and to some extent ALTAS and CMS) are primed to conduct searches for $B$ mesons decaying into baryons and missing energy. Doing so could unveil the nature of baryogenesis and therefore our very existence.

\newpage
\section{Outlook}
\label{sec:conc}
In this work we have presented the results of a LCSR calculation of the form factors and branching fractions for the seemingly baryon number violating decays of  $B_{s,d}^0$ and $B^+$ mesons into a dark sector baryon $\psiB$ and various SM baryon final states. Figures \ref{fig:Br_light} and \ref{fig:Br_heavy} show the results for the maximum possible branching fractions (in light of LHC constraints on the Wilson coefficients) and associated uncertainties for each of the four flavorful variation on the operators that could produce the baryon asymmetry. These results will serve to guide experimental searches for these exclusive decay modes. Several interesting future directions exist, which we now comment briefly upon.    

In the present work we considered operators involving only right handed fermions as would arise in the model defined by the Lagrangian in Eq.~\eqref{eq:Lmodel} i.e. where the colored mediator is a scalar and has SM hyper-charge assignment $-1/3$. However, other UV completions do exist corresponding to different charge assignments for the mediator (see the models in~\cite{Alonso-Alvarez:2021qfd}). For instance, the mediator could be a doublet under $SU(2)_L$ which would give rise to four fermion operators containing left handed quarks. The branching fractions arising from these operators would lead to non-vanishing contributions from the $F_L^{\mathcal{O}}(q^2)$ and $\widetilde{F}_R^{\mathcal{O}}(q^2)$ form factors. The computation of these form factors is a trivial extension and we leave it to future work. 

The model of Eq.~\eqref{eq:Lmodel} would also give rise to the decay of a baryon
into light mesons and missing energy e.g. $\Lambda_b \rightarrow
\bar{\psi}_\mathcal{B} + \pi^0$ through the $\mathcal{O}_{b,ud}$ and $\mathcal{O}_{d,ub}$ (see~\cite{Elor:2018twp} for other decay modes). While the branching fraction of such decays do not directly feed into the Boltzmann equations that track the generated baryon asymmetry, they do serve as an indirect probe of $B$-Mesogenesis. Therefore, it would be interesting for experimental searches to target these decays as well. Indeed, the proposed search at LHCb~\cite{Rodriguez:2021urv} would be capable of measuring these $b$-flavored baryon decays. As such it is also a worthy peruse to also compute the predicted branching fractions for these decays. To do so one can once again use the LCSR machinery. The starting point would be a three point correlator and we leave the details of this interesting technical pursuit to future work. 

Another application of the LCSR technique would to calculate the branching fractions of the $B$ meson decay's in Table~\ref{tab:decays} where multiple final state light mesons are radiated e.g. $\pi^0$.

\acknowledgments
We thank Zoltan Ligeti and Stefano Rigolin for useful conversations. 
G.E. is supported by the Cluster of Excellence {\em Precision Physics, Fundamental Interactions and Structure of Matter\/} (PRISMA${}^+$ -- EXC~2118/1) within the German Excellence Strategy (project ID 39083149).
A.W.M.G. is supported by the European Union’s Horizon 2020 research and innovation programme under the Marie Sklodowska-Curie grant agreement 860881 (HIDDEN).

\appendix

\newpage
\section{Technical Details}
We now give an overview of some of the more technical details of our derivation using the LCSR, the DAs for decays to heavy charmed final state baryons, and finally numerical results for extracting form factors. 
\subsection{Form Factors from LCSR}
\label{sec:FF_LCSR}

The central object of this approach is the correlation function mediating the $B\to \mathcal{B}_\mathrm{SM}+\psi_\mathcal{B}$ transition:
%
\begin{equation}
  \Pi(P,q)=i \int d^4x e^{i(P+q)\cdot x} \bra{0}\mathrm{T}\{j_B(x),\mathcal{O}(0)\}\ket{\mathcal{B}_\mathrm{SM}},  
  \label{eq:correlation_function}
\end{equation}
%
where $j_B=m_b\, \bar{b}i \gamma^5 u$ is the $B$ meson interpolating current carrying four-momentum $(P+q)$, $P$ and $q$ are the $\mathcal{B}$ and $\psi_\mathcal{B}$
momenta respectively while $\mathcal{O}$ is one of the operators defined in Eqs.~(\ref{eq:OpsDef}).
The correlation function in Eq.~(\ref{eq:correlation_function})
can be rewritten by employing an hadronic dispersion relation in the
$(P+q)^2$ variable with the $B$-meson pole isolated 
%
\begin{equation}
  \Pi(P,q)=\frac{m_B^2f_B\bra{B(P+q)}
  \mathcal{O}\ket{\mathcal{B}_\mathrm{SM}(P)}}{m_B^2-(P+q)^2}+\int_{s_h}^\infty \!\!\! ds\,\, \frac{\rho(s,q^2)}{s-(P+q)^2},\label{eq:dispersion_relation}
\end{equation}
%
where we used the identity $\bra{0}j_B\ket{B}=m_B^2f_B$. Here $\rho$ is the spectral density encapsulating contributions from the excited and continuum states above threshold. 
$\Pi(P,q)$ can be decomposed in terms of Lorentz covariant structures: 
%
\bea
   && \Pi(P,q)=\Pi_R((P+q)^2,q^2)\bar{u}_{\mathcal{B}_R}+\Pi_L((P+q)^2,q^2)\bar{u}_{\mathcal{B}_L}\nonumber\\
    &&\qquad\qquad\qquad +\widetilde{\Pi}_R((P+q)^2,q^2)\slashed{q}\bar{u}_{\mathcal{B}_R}
            +\widetilde{\Pi}_L((P+q)^2,q^2)\slashed{q}\bar{u}_{\mathcal{B}_L}, 
            \label{eq:correlation_decomposition}
\eea
%
where the $\Pi_{L,R}$, $\widetilde{\Pi}_{L,R}$ are Lorentz invariant functions.
Substituting Eq.~(\ref{eq:form_factors_decomposition}) and Eq.~(\ref{eq:correlation_decomposition}) 
in Eq.~(\ref{eq:dispersion_relation})
exposes the form factors in the hadronic matrix element in terms of the Lorentz invariant
$\Pi_I$\footnote{The superindex $I$ is intended to include
$\Pi_{L,R}$ and $\widetilde{\Pi}_{L,R}$.} functions, namely 
%
\bea
   && \Pi_I((P+q)^2,q^2)=\frac{m_B^2f_B F_I(q^2)}{m_B^2-(P+q)^2}+\int_{s_h}^\infty \!\!\! ds\,\, \frac{\rho_I(s,q^2)}{s-(P+q)^2} \,.\label{eq:invariant_dispersion_relation}
\eea
%
Note that to compute $\Pi_I=\widetilde{\Pi}_{R,L}$ requires the substitution $F_I\to \mbaryo^{-1} \widetilde{F}_{R,L}$. 
In the region where the momenta are very far off-shell,
$(P+q)^2,q^2\ll m_b^2$, the integral Eq.~(\ref{eq:correlation_function}) is dominated by modes near the light-cone  $x^2=0$. In this case the 
 computation is
carried out via a light-cone OPE convoluted with the 
distribution amplitudes (DAs) of the baryons~\cite{RQCD:2019hps},  we refer to the correlation function computed in this way as $\Pi_I^\mathrm{OPE}$. 
The invariant amplitudes calculated in this way can be written with a a convenient 
dispersion relations, i.e.
%
\bea
   && \Pi_I^\mathrm{OPE}((P+q)^2,q^2)=\frac{1}{\pi}\int_{m_b^2}^\infty\!\!\! ds\,\, \frac{\Im{\Pi_I^\mathrm{OPE}(s,q^2)}}{s-(P+q)^2}.\label{eq:OPE_invariant}
\eea
%
Substituting Eq.~(\ref{eq:OPE_invariant}) and evoking semi-global quark hadron duality we have
%
\beq
    \int_{s_h}^\infty ds \frac{\rho_I(s,q^2)}{s-(P+q)^2}=\frac{1}{\pi}\int_{s_0^B}^\infty \!\!\! ds\,\, \frac{\Im{\Pi_I^\mathrm{OPE}(s,q^2)}}{s-(P+q)^2},
\eeq
%
in Eq.~(\ref{eq:invariant_dispersion_relation}). We then take the standard Borel transformation, $(P+q)^2\to M^2$,
to obtain the LCSR master formula for the form factors:
\begin{equation}
\label{eq:LCSRmasterformula}
    m_{B}^2f_{B} F_I(q^2)e^{-m^2_{B}/M^2}=\frac{1}{\pi} \int_{m_b^2}^{s_0^B}\!\!\! ds\,\, e^{-s/M^2}\Im{\Pi_I^\mathrm{OPE}(s,q^2)}\,,
\end{equation}
where $s_0^B$ is the effective
threshold.
Once again, whenever one wishes to consider $\Pi_I=\widetilde{\Pi}_{R,L}$ the correct substitution for 
the form factor is simply $F_I\to \mbaryo^{-1} \widetilde{F}_{R,L}$. 
Now, to derive Eqs.~(\ref{eq:R1}--\ref{eq:L1}) from Eq.~\eqref{eq:LCSRmasterformula} one 
simply has to revert back to the original integration variable $\alpha$, and  substitute the invariant amplitude calculated at leading twist in the OPE. Note that whenever the momentum combination $(P+q)^2$ appears in the numerator of the invariant amplitudes one has to remove it via a subtraction of a term  independent of $(P+q)^2$. These can than be omitted as they vanish after the Borel transformation. 

\subsection{Charmed Sector Distribution Amplitudes}
\label{sec:charmed}

Since there are no Lattice QCD results for the DAs beyond the baryon octet, one
is forced to employ a model to tackle heavier states.
To extended the analysis to charmed baryons one can employ 
the  DAs obtained in the HQET~\cite{PhysRevD.55.272,Ali:2012pn} and assume 
heavy quark symmetry~\cite{Neubert:1993mb}. 
The only resonances participating in the decays of Table~\ref{tab:decays} 
are of the $SU(3)_F$ anti-triplet and sextet with spin parity $J^P=1/2^+$.
Let us first consider the anti-triplet states, characterized by $j^P=0^+$ for the light di-quark system\footnote{$j$ is the spin of the valence light quark couple.}, these can be 
decomposed at leading twist as
%
\beq
    \langle 0| q_{1\alpha}(t_1)q_{2\beta}(t_2)Q_\gamma(0)| H_Q^{j=0}\rangle=
    \left(\sum_i B^{j=0}_i \left[ \Gamma_iC^{-1} \right]_{\alpha\beta}\right)(u_{H_Q})_\gamma \,,
    \label{eq:decomposition3}
\eeq
%
where the basis is defined as $\Gamma_1=\gamma^5\slashed{\bar{n}},$ 
$\Gamma_2=1/2\sigma_{\bar{n}n}\gamma^5,$ $\Gamma_3=\gamma^5$ and
$\Gamma_4=\gamma^5\slashed{n},$ and $(u_{H_Q})_\gamma$ 
is the spinor associated to the heavy baryon.
Here $n_\mu$ is the light cone 
vector along which the light-quarks are aligned, $\bar{n}_\mu$
is its orthogonal direction and $\sigma_{\bar{n}n}=\sigma_{\mu\nu}\bar{n}^\mu n^\nu$. To find the coefficients of the 
decomposition, one imposes Eq.~(23) of Ref.~\cite{Ali:2012pn}.
At leading-twist only $B^{j=0}_1$ is non-zero,
\beq
   B^{j=0}_1=\frac{1}{8}v^+\psi^2(t_1,t_2)f_{H^{j=0}_Q}\,,
   \label{eq:decompositionA}
\eeq
where $v^+$ is the projection of the velocity of the heavy quark
$Q_\gamma$ along the $n_\mu$ direction. $\psi^2$ and $f_{H^{j=0}_Q}$ are defined in~\cite{Ali:2012pn}.
In the $j=1^+$ case a similar decomposition holds~\cite{Aliev:2022noh}
with the substitutions $\Gamma_1=\slashed{n},$ 
$ (u_{H_Q})_\gamma\to (\gamma^5 u_{H_Q})_\gamma$, and
\beq
B^{j=1}_1= \frac{1}{8v^+}\psi^2(t_1,t_2)f_{H^{j=1}_Q} \,.
\label{eq:decomposition6}
\eeq
Now one can connect the DAs in~\cite{Ali:2012pn} with the ones defined 
in~\cite{RQCD:2019hps}. 
One can see that, in the $j=0$, case the equality connecting the two definitions
is
\beq
    \frac{f_{H_Q^{j=0}}}{M_Q}\psi^2(x,0)=\mathcal{A}(x,0,0)\,,
\eeq
this can be proved by noting that $M_Q v^+ \slashed{n}=\slashed{\tilde{n}},$
where $\slashed{\tilde{n}}$ is defined in Eq.~(4) of Ref.~\cite{RQCD:2019hps}.
For practical reasons it is useful to find an identity between the Fourier 
transformed of the two  i.e. 
\bea
   \int d\alpha_1 d\alpha_2 e^{-i\, P\cdot x \,\alpha_1}  &&\delta\left(\frac{M_Q-m_Q}{M_Q}-\alpha_1-\alpha_2\right) A\left(\alpha_1,\alpha_2,\frac{m_Q}{M_Q}\right)\nonumber\\
   &&\qquad\quad=\int d\alpha_1^\prime d\alpha_2^\prime e^{-i\, P\cdot x \,\alpha_1^\prime}\delta\left(1-\alpha_1^\prime-\alpha_2^\prime\right) \phi(\alpha_1^\prime,\alpha_2^\prime),\label{eq:dictionary}
\eea
where $M_Q$ and $m_Q$ are the masses of the baryon and the 
heavy quark respectively and $\phi$ is defined as 
\beq
\phi(\alpha_1,\alpha_2)=f_{H^j_Q} M^3_Q \left(1-\frac{m_Q}{M_Q}\right)^4\left[\alpha_1\alpha_2\sum_{n=0}^{2}\frac{a_n}{\epsilon_n^4}\frac{C^{3/2}_n(\alpha1-\alpha_2)}{|C^{3/2}_n|}e^{-(M_Q-m_Q)/\epsilon_n}\right].\label{eq:heavy_distribution}
\eeq
The function $\mathcal{A}$ is defined  as
as 
\beq
    \mathcal{A}(x,0,0)=\int\! d\alpha_1 d\alpha_2 d\alpha_3\,\, \delta(1-\sum_i \alpha_i)\, e^{-i\, P\cdot x \,\alpha_1}\,A\left(\alpha_1,\alpha_2,\alpha_3\right)\,,
\eeq
and the HQET hypothesis introduces the constraint $\delta(\alpha_3-m_Q/M_Q)$ in Eq.~(\ref{eq:dictionary}).
A similar argument can be used for the $j=1$ case, see~\cite{Aliev:2022noh}, by swapping $A$ for the $V$ function.
These DAs for charmed baryons can then be substituted in the results of Eqs.~(\ref{eq:R1}--\ref{eq:L1}) and, after the fit discussed in App.~\ref{app:Numeric}, will produce the results of Figure \ref{fig:Br_heavy}.


\subsection{BCL Fits and Parameters}
\label{app:Numeric}

\begin{table}
\center
\begin{tabular}{| c | c |}
\hline
                       &                   \\
    Parameter          &   Numerics        \\    
    \hline
Renormalization scale  &    $\mu= 3$ GeV    \\ 
\hline
Effective threshold   &    $s_0^B= 39\pm 1.25$ GeV$^2$    \\ 
\hline
Borel Parameter Squared   & $M^2= 16 \pm 4$ GeV$^2$    \\ 
\hline
\end{tabular}
\caption{Values and intervals of the internal parameters.\label{tab:values}}
\end{table}

\begin{table}
\begin{tabular}{| c | c || c | c | c |}
\hline
            Decay   &        Fit    &        &                                       &                \\
            Channels &  Parameters  & $R^b_I$ &     $R^{d_k}_I$                     &  $\widetilde{L}_1$  \\
            \hline\hline
$B_d\!\to  \psi_\mathcal{B}+\,n$ &$F_I(0)$       & $1.94_{\pm 0.12}\!\cdot\! 10^{-2}$&$1.94_{\pm 0.12}\!\cdot\! 10^{-2}$ & n.a. \\
                                  & $b_I$        & $3.07^{\,\,0.11}_{\,\,0.17}$    &$3.07^{\,\,0.11}_{\,\,0.17}$ &      n.a.       \\
\hline
$B_s\!\to  \psi_\mathcal{B}+\,\Lambda$ &$F_I(0)$     &   n.a.    &     n.a.   &        n.a.                 \\
                                        & $b_I$     &   n.a.       &        n.a.     &         n.a.         \\
\hline
$B^+\!\to \psi_\mathcal{B} +\,p $ &$F_I(0)$ & $-9.20_{\pm 0.29}\!\cdot\! 10^{-3}$&$1.94_{\pm 0.11}\!\cdot\! 10^{-2}$   & $5.87_{\pm 0.29}\!\cdot\! 10^{-4}$  \\
                                  & $b_I$   & $-1.48^{\,\,0.22}_{\,\,0.21}$     &$3.07^{\,\,0.11}_{\,\,0.17}$        & $-2.83^{\,\,0.26}_{\,\,0.29}$     \\
\hline
\hline
$B_d\!\to \psi_\mathcal{B}+\,\Lambda $ &$F_I(0)$ & $4.09_{\pm 0.11}\!\cdot\! 10^{-2}$   &$4.09_{\pm 0.11}\!\cdot\! 10^{-2}$ & n.a.\\
                                         & $b_I$   & $6.91^{\,\,0.18}_{\,\,0.18}$       & $6.91^{\,\,0.18}_{\,\,0.18}$   &   n.a.         \\
\hline
$B_s\!\to \psi_\mathcal{B}+\,\Xi^0 $ &$F_I(0)$ & $-1.83_{\pm 0.09}\!\cdot\! 10^{-2}$& $4.69_{\pm 0.19}\!\cdot\! 10^{-2}$& $9.54_{\pm 0.65}\!\cdot\! 10^{-4}$ \\
                              & $b_I$       & $-1.83^{\,\,-0.17}_{\,\,-0.04}$     &$1.57^{\,\,0.23}_{\,\,0.25}$     &   $-3.17^{\,\,0.39}_{\,\,0.45}$     \\
\hline
$B^+\!\to \psi_\mathcal{B}+\,\Sigma^+ $&$F_I(0)$&$-1.73_{\pm 0.04}\!\cdot\! 10^{-2}$&$3.12_{\pm 0.14}\!\cdot\!10^{-2}$ & $9.29_{\pm 0.54}\!\cdot\!10^{-4}$ \\
                            & $b_I$       & $-1.07^{\,\,-0.12}_{\,\,-0.11}$&$4.63^{\,\,-0.02}_{\,\,-0.02}$          &    $-3.00^{\,\,0.29}_{\,\,0.32}$    \\
\hline
\hline
$B_d\!\to \psi_\mathcal{B}+\,\Sigma^0_c $&$F_I(0)$&$7.19_{\pm 1.8}\!\cdot\! 10^{-3}$& $3.76_{\pm 1.1}\!\cdot\!10^{-2}$&  $8.94_{\pm 2.3}\!\cdot\!10^{-3}$ \\
                                 & $b_I$     & $2.49^{\,\,-0.91}_{\,\,-1.50}$        & $13.0^{\,\,-1.8}_{\,\,-3.2}$    &    $1.61^{\,\,-1.2}_{\,\,-2.0}$  \\
\hline
$B_s\!\to \psi_\mathcal{B}+\,\Xi^0_c $&$F_I(0)$&$6.92_{\pm 1.7}\!\cdot\! 10^{-3}$& $3.53_{\pm 0.94}\!\cdot\!10^{-2}$  & $8.56_{\pm 2.2}\!\cdot\!10^{-3}$\\
                              & $b_I$     & $2.70^{\,\,-0.89}_{\,\,-1.40}$        &$13.5^{\,\,-1.8}_{\,\,-3.1}$    &   $1.81^{\,\,-1.2}_{\,\,-1.9}$    \\
\hline
$B^+\!\to \psi_\mathcal{B}+\,\Sigma^+_c $&$F_I(0)$&$7.18_{\pm 1.80}\!\cdot\! 10^{-3}$&$3.76_{\pm 1.10}\!\cdot\!10^{-2}$&  $8.94_{\pm 2.30}\!\cdot\!10^{-3}$\\
                                  & $b_I$         & $2.51^{\,\,-0.92}_{\,\,-1.50}$   &$13.0^{\,\,-1.8}_{\,\,-3.2}$     &  $1.63^{\,\,-1.20}_{\,\,-2.00}$    \\
\hline
\hline
$B_d\!\to \psi_\mathcal{B}+\,\Xi^0_c     $&$F_I(0)$&$8.17_{\pm 1.90}\!\cdot\! 10^{-3}$& $4.16_{\pm 1.10}\!\cdot\!10^{-2}$ &$1.01_{\pm 0.25}\!\cdot\!10^{-2}$ \\
                                          & $b_I$  & $2.64^{\,\,-0.85}_{\,\,-1.40}$   & $13.2^{\,\,-1.7}_{\,\,-3.0}$ &   $1.77^{\,\,-1.10}_{\,\,-1.90}$     \\
\hline
$B_s\!\to \psi_\mathcal{B}+\,\Omega_c     $&$F_I(0)$&$1.24_{\pm 0.25}\!\cdot\! 10^{-2}$& $5.12_{\pm 1.20}\!\cdot\!10^{-2}$ & $1.51_{\pm 0.32}\!\cdot\!10^{-2}$\\
                                & $b_I$  & $3.90^{\,\,-1.05}_{\,\,-1.56}$                &$16.5^{\,\,-2.10}_{\,\,-3.30}$      &  $3.33^{\,\,-1.35}_{\,\,-2.07}$   \\
\hline
$B^+\!\to \psi_\mathcal{B}+\,\Xi^+_c     $&$F_I(0)$&$8.17_{\pm 1.90}\!\cdot\! 10^{-3}$& $4.18_{\pm 1.11}\!\cdot\!10^{-2}$& $1.01_{\pm 0.32}\!\cdot\!10^{-2}$ \\
                                         & $b_I$  & $2.61^{\,\,-0.84}_{\,\,-1.37}$   &$13.2^{\,\,-1.7}_{\,\,-2.9}$  &    $1.73^{\,\,-1.40}_{\,\,-2.70}$   \\
\hline

\end{tabular}
\caption{Fit parameters for the different decay channels.
\label{tab:fit}}
\end{table}

Here we will first discuss the values of the parameters we 
used to obtain the results of Sec.~\ref{sec:results}, we then  present the details regarding the fit of the form factors in the kinematically allowed region.

The DAs are taken from the latest Lattice QCD results for the octet baryons~\cite{RQCD:2019hps}, 
while the HQET Light-Cone DAs can be found in~\cite{Ali:2012pn}.
Finally the wave function normalizations at the origin for the charmed baryons are then taken from~\cite{Aliev:2018hre,PhysRevD.56.3943,Agaev:2017lip}.
DAs are expanded in terms of orthogonal polynomials in such a way that the  one loop scale dependence of their coefficients is autonomous (conformal partial wave expansion). In this way the non-perturbative information is encoded 
in scale dependent coefficients known as \emph{shape parameters}. The running effects for the shape parameters and the couplings are taken
into account using the results in~\cite{Anikin:2013aka,Ball:2008fw}.
The Borel parameter $M^2$ and the renormalization scale $\mu$
are taken in the optimal ranges for a correlation function with the $B$-meson 
interpolation current~\cite{Khodjamirian:2022vta,Khodjamirian:2017fxg,Khodjamirian:2020mlb}. 
The effective threshold $s_0^B$ is then fixed by fitting the mass of the
$B$-meson obtained from the logarithmic derivative of the sum rule with 
the measured one, at each value of $M^2$,
by allowing a variation of $3\%$~\cite{Khodjamirian:2011ub,Duplancic:2008ix}.
Masses and parameters of mesons and quarks are taken from~\cite{Workman:2022ynf},
while the internal values used to reproduce the results are reported in Table~\ref{tab:values}.

The form factors in Eq.~(\ref{eq:R1}--\ref{eq:L1}) are calculated for $q^2\ll m_b^2$,
including the spacelike region $q^2<0$.
To extrapolate the result in the whole kinematic region it is customary
to employ the z-expansion, this involves mapping $q^2$ to a new variable $z$ defined as 
\beq
    z(q^2)=\frac{\sqrt{t_+-q^2}-\sqrt{t_+-t_0}}{\sqrt{t_+-q^2}-\sqrt{t_+-t_0}}\,,
\eeq
where $t_+=(m_B+\mbaryo)^2$ and $t_0=(m_B+\mbaryo)(\sqrt{m_B}-\sqrt{\mbaryo})$. This
change of variables maps the whole complex $q^2$ plane onto the unit disk in the $z$
plane, with the paths along the branch cut mapped on the circle enclosing the unit disk. 
Moreover the choice of $t_0$ makes it so that
the kinematically allowed region is centered around the origin and limited 
by $|z|<0.08$ for all the decays considered. We choose to fit our results onto the
Bourrely-Caprini-Lellouch (BCL) expansion~\cite{Bourrely:2008za} slightly modified 
according to~\cite{Khodjamirian:2017fxg},
\beq
    F_I(q^2)=\frac{F_I(0)}{1-q^2/m^2_{\Lambda_b}}\left\{ 1+ b_I\left[z(q^2)-z(0)
                        +\frac{1}{2}\left(z(q^2)^2-z(0)^2\right) \right]\right\}\,.\label{eq:BCL}
\eeq
 We use the interval $-5.0<q^2<1$ GeV$^2$ to perform the fit. 
The uncertainties are obtained by varying the internal parameters in their range and 
adding them in quadrature with the errors on the shape parameters of the DAs.
In the case of charmed baryons the violation of heavy quark symmetry is obtained
by adding a flat 15\% error~\cite{Khodjamirian:2017zvq} on the couplings and on the parameters of the DAs.
A report on the two parameter fit for each operator and channel is included in Table~\ref{tab:fit}.
For each channel we report the values of $F_I(0)$ and $b$ for the three possible form factors. 
In the first two columns $R_I$ from $\mathcal{O}_{b,u_i,d_j}$ and $R_I$ from $\mathcal{O}_{d_k,u_i,b}$ are reported. They are dubbed $R_I^b$ and $R_I^{d_k}$ respectively.
Finally the last column is dedicated to $\widetilde{L}_1$. Note that $\widetilde{L}_1$ does not depend on the operator.

The values reported in Table~\ref{tab:fit} define three distinct curves for each form factor of 
a single transition corresponding to the extremes of the region allowed by the errors and to
its central value. 
The latter is recovered by substituting the entries reported in Table~\ref{tab:fit} in Eq.(\ref{eq:BCL}).
The other two curves correspond to the values of the parameters $F_I^\pm(0) = F_I(0)\pm \delta (F_I(0))$ and
$b^+_I=b_I+b^\mathrm{Sup}_I$ and $b^-_I=b_I-b_{I\mathrm{Sub}}$, where $\delta (F_I(0))$ 
is the error reported in Tab \ref{tab:fit} while $b^\mathrm{Sup}_I$ and $b_{I\mathrm{Sub}}$ are 
the superscript and subscript corresponding to the parameter $b$ respectively.

\bibliographystyle{JHEP}
\bibliography{ref}

\end{document}